\documentclass[aps,prb,twocolumn,showpacs,preprintnumbers,amsmath,amssymb,superscriptaddress]{revtex4}%

\usepackage{graphicx}
\usepackage{dcolumn}
\usepackage{bm}
\usepackage{color}
\begin{document}


\title{Electrostatically tuned quantum superconductor-metal-insulator transition at the LaAlO$_{3}$/SrTiO$_{3}$\
interface}

\author{T. Schneider}
\email{tschnei@physik.unizh.ch}
\affiliation{Physik-Institut der Universit\"{a}t Z\"{u}rich, Winterthurerstrasse 190, CH-8057, Switzerland}
\author{A.D. Caviglia}
\author{S. Gariglio}
\author{N. Reyren}
\author{J.-M. Triscone}
\affiliation{DPMC, University of Geneva, 24 Quai Ernest-Ansermet, 1211 Geneva 4,Switzerland.}

\begin{abstract}
Recently superconductivity at the interface between the insulators LaAlO$_{3} $ and SrTiO$_{3}$ has been tuned with the electric field effect to an unprecedented range of transition temperatures. Here we perform a detailed
finite size scaling analysis to explore the compatibility of the phase
transition line with Berezinskii-Kosterlitz-Thouless (BKT) behavior and a
2D-quantum phase(QP)-transition. In an intermediate regime, limited by a
gate voltage dependent limiting length, we uncover remarkable consistency
with a BKT-critical line ending at a metallic quantum critical point,
separating a weakly localized insulator from the superconducting phase. Our
estimates for the critical exponents of the 2D-QP-transition, $z\simeq 1$
and $\overline{\nu }\simeq 2/3$, suggest that it belongs to the 3D-xy universality class.
\end{abstract}

\pacs{74.78.-w, 74.40.+k, 74.90.+n, 74.78.Fk}
\maketitle

\section{INTRODUCTION}
At the interface between oxides, electronic properties have been generated,
different from those of the constituent materials.\cite{ohtomo,reyren,bousquet} In particular, the interface between LaAlO$_{3}$ and SrTiO$_{3}$, two excellent band insulators, found to be conducting in
2004 \cite{ohtomo} attracted a lot of attention \cite{okamoto,brinkman,willmott,siemons,herranz,pauli}. Recently, different
ground states, superconducting and ferromagnetic, have been reported for
this fascinating system.\cite{reyren} In a recent report \cite{caviglia},
it was shown that the electric field effect can be used to map the phase
diagram of this interface system revealing, depending on the doping level, a
superconducting and non-superconducting ground state and evidence for a
quantum phase transition.

 Continuous quantum phase transitions are transitions at absolute zero in
which the ground state of a system is changed by varying a parameter of the
Hamiltonian.\cite{sondhi,book,tsben} The transitions between superconducting and
insulating behavior in two-dimensional systems tuned by disorder,
film thickness, magnetic field or with the electrostatic field effect are
believed to be such transitions.\cite{book,tsben,goldman,goldman2,parendo,aubin,parendo2,matthey}

 Here we present a detailed finite size scaling analysis of the temperature
and gate voltage dependent resistivity data of Caviglia \textit{et al}.\cite{caviglia} to explore in the LaAlO$_{3}$/SrTiO$_{3}$\ system the nature of
the phase transition line and of its endpoint, separating the
superconducting from the insulating ground state. For this purpose we
explore the compatibility of the normal state to superconductor transition
with Berezinskii-Kosterlitz-Thouless (BKT) critical behavior.\cite{berz,kosterlitz} Our analysis of the temperature dependence of the sheet resistance at various fixed gate voltages uncovers a rounded BKT-transition.
The rounding turns out to be fully consistent with a standard finite size
effect whereupon the correlation length is prevented to grow beyond a
limiting length $L$. Indeed, a finite extent of the homogeneous domains will prevent
the correlation or localization length to grow beyond a limiting length $L$
and, as a result, a finite size effect occurs. Because the correlation
length does not exhibit the usual and relatively slow algebraic divergence
as $T_{c}$ is approached, the BKT-transition is particularly susceptible to
such finite size effects. Nevertheless, for sufficiently large $L $ the
critical regime can be attained and a finite size scaling analysis provides
good approximations for the limit of fundamental interest, $L\rightarrow
\infty $.\cite{book,fisherm,privman}

 As will be shown below, our finite size scaling analysis uncovers close to
the QP-transition a gate voltage dependent limiting length. According to
this electrostatic tuning does not change the carrier density only but the
inhomogeneity landscape as well. The finite size scaling analysis also allows us to determine the gate voltage dependence of the BKT-transition temperature $T_{c}$, of the associated fictitious infinite system. This critical line, $T_{c}$ versus gate voltage, ends at a quantum critical point at the gate
voltage $V_{gc}$. Here the sheet conductivity tends to $\sigma _{\square
}\left( T=0,V_{gc}\right) \simeq 2.52\cdot 10^{-4}$ ($\Omega ^{-1}$) which
is comparable to the quantum unit of conductivity $4e^{2}/h\simeq 1.55\cdot
10^{-4}$ ($\Omega ^{-1}$) for electron pairs, emphasizing the importance of
quantum effects. Its limiting $T^{2}$ temperature dependence points to Fermi
liquid behavior at quantum criticality. The estimates for the critical
exponents of the 2D-QP-transition, $z\simeq 1$ and $\overline{\nu }\simeq 2/3
$, suggest that it belongs to the 3D-xy universality class. In the normal
state we observe non-Drude behavior, consistent with the evidence for weak
localization. To identify the nature of the insulating phase from the temperature
dependence of the resistance, we perform a finite size scaling analysis,
revealing that the growth of the diverging length associated with weak
localization is limited and gate voltage dependent as well. Nevertheless, we
observe in both, the temperature and magnetic field dependence of the
resistance, the characteristic weak localization behavior, pointing to
a renormalized Fermi liquid. In addition we explore the $T_{c}$
dependence of the vortex core radius and the vortex energy. These properties
appear to be basic ingredients to understand the variation of $T_{c}$. In the superconducting phase we observe consistency with the standard quantum scaling form for the resistance, while in the weakly localized phase
it appears to fail. In contrast to the quantum scaling approach we obtain the scaling function
in the superconducting phase explicitly. It is controlled
by the BKT-phase transition line and the vortex energy.

 In Section II we sketch the theoretical background and present the detailed
analysis of the resistivity data of Caviglia \textit{et al}.\cite{caviglia}
We close with a brief summary and some discussion.

\section{THEORETICAL BACKGROUND AND DATA ANALYSIS}

\subsection{BKT-TRANSITION}

To explore the compatibility with BKT critical behavior we invoke the
characteristic temperature dependence of the correlation length above $T_{c}$,\cite{kosterlitz}
\begin{equation}
\xi \left( T\right) =\xi _{0}\exp \left( 2\pi /\left( bt^{1/2}\right)
\right) ,t=\left\vert T/T_{c}-1\right\vert ,
\label{eq1}
\end{equation}
where $\xi _{0}$ is the classical vortex core radius and $b$ is related to
the energy needed to create a vortex.\cite{ambek,finotello,steel,tilley} Note
that $b$ also enters the temperature dependence of the magnetic penetration
depth $\lambda $ below the universal Nelson-Kosterlitz jump \cite{ambek}:
\begin{equation}
\lambda ^{2}\left( T_{c}\right) /\lambda ^{2}\left( T\right) =\left(
1+b\left\vert t\right\vert ^{1/2}/4\right) .
\label{eq2}
\end{equation}
Moreover, $b$ is related to the vortex energy $E_{c}$ in terms
of\cite{steel,dahm}
\begin{equation}
b=f\left( E_{c}/\left( k_{B}T_{c}\right) \right) .
\label{eq3}
\end{equation}
Invoking dynamic scaling the resistance $R$ scales in $D=2$ as \cite{book}
\begin{equation}
R\propto \xi ^{-z_{cl}},
\label{eq4}
\end{equation}
where $z_{cl}$ is the dynamic critical exponent of the classical dynamics. $z_{cl}$ is usually not questioned to be anything but the value that describes simple diffusion: $z_{cl}=2$.\cite{pierson} Combining these
scaling forms we obtain
\begin{equation}
\frac{R\left( T\right) }{R_{0}}=\left( \frac{\xi _{0}}{\xi \left( T\right) }
\right) ^{2}=\exp \left( {-b_{R}}(T-T_{c})^{-1/2}\right) ,
\label{eq5}
\end{equation}
with
\begin{equation}
b_{R}=4\pi T_{c}^{1/2}/b,\text{ }R_{0}\propto 1/\xi _{0}^{2}.
\label{eq6}
\end{equation}
Accordingly the compatibility of experimental resistivity data with the
characteristic BKT-behavior can be explored in terms of
\begin{equation}
\left( d\ln R/dT\right) ^{-2/3}=\left( 2/b_{R}\right) ^{2/3}(T-T_{c}).
\label{eq7}
\end{equation}
Because the correlation length does not exhibit the usual and relatively
slow algebraic divergence as $T_{c}$ is approached (Eq. (\ref{eq1})) the
BKT-transition is particularly susceptible to the finite size effect. It
prevents the correlation length to grow beyond a limiting lateral length $L$
and leads to a rounded BKT-transition. Nevertheless, for sufficiently large $L $ the critical regime can be attained and a finite size scaling analysis
allows good approximations to be obtained for the limit $L\rightarrow \infty
$\cite{book,fisherm} including estimates for $T_{c}$, $b_{R}$, $R_{0}$,
and their gate voltage dependence. In the present case potential candidates
for a limiting length include the finite extent of the homogenous regions
and the failure to cool the electron gas down to the lowest temperatures. In
the latter case $\ L$ is given by the value of the correlation length at the
temperature where the failure of cooling sets in. In any case finite size
scaling predicts that $R\left( T,L\right) $
adopts the form
\begin{eqnarray}
\frac{R\left( T,L\right) }{R\left( T,\infty \right) } &=&\left( \frac{\xi
\left( T,0\right) }{\xi \left( T,\infty \right) }\right) ^{2}=g(x)  \nonumber
\\
&=&\frac{R\left( T,L\right) }{R_{0}}\text{exp}\left( b_{R}\left\vert
T-T_{c}\right\vert ^{-1/2}\right) ,
\label{eq8}
\end{eqnarray}
where
\begin{equation}
x=\frac{\text{exp}\left( b_{R}\left\vert T-T_{c}\right\vert ^{-1/2}\right) }{%
R_{0}L^{2}}\propto \left( \frac{\xi \left( T,\infty \right) }{L}\right) ^{2}.
\label{eq8a}
\end{equation}
$g(x)$ is the finite size scaling function. If $\xi \left( T,\infty \right)
<L$ critical behavior can be observed as long as $g(x)\simeq 1$, while for $%
\xi \left( T,\infty \right) >L$ the scaling function approaches $g(x)\propto
x$ so $R(T)$exp$\left( b_{R}\left\vert T-T_{c}\right\vert ^{-1/2}\right)
/R_{0}$ tends to $\left( g/R_{0}\right) $exp$\left( b_{R}\left\vert
T-T_{c}\right\vert ^{-1/2}\right) $ with $g\simeq 1/L^{2}$.

We are now prepared to explore the evidence for BKT-behavior. In Fig. \ref%
{fig1} we show \ $\left( d\ln R/dT\right) ^{-2/3}$ \textit{vs}. $T$ for $%
V_{g}=40$ V. In spite of the rounded transition there is an intermediate
regime revealing the characteristic BKT-behavior (\ref{eq7}), allowing us to
estimate $R_{0}$, $b_{R}$ and $T_{c}$. As can be seen in the inset of \ref%
{fig1}, depicting $R(T)$ exp$\left( b_{R}\left\vert T-T_{c}\right\vert
^{-1/2}\right) /R_{0}$ \textit{vs}. $\left( 1/R_{0}\right) $exp$\left(
b_{R}\left\vert T-T_{c}\right\vert ^{-1/2}\right) $, the rounding of the
transition is remarkably consistent with a standard finite size effect. The
horizontal line corresponds to $\xi <L$ where critical behavior can be
observed as long as $g(x)\simeq 1$, while the dashed one characterizes the
rounded regime where $\xi >L$. Here the scaling function approaches $%
g(x)\propto x$ and $R(T,L)$ tends to $g\propto $ $L^{-2}$. Independent
evidence for BKT-behavior was also established in earlier work in terms of
the current-voltage characteristics.\cite{reyren}

\begin{figure}[htb]
\includegraphics[width=1.0\linewidth]{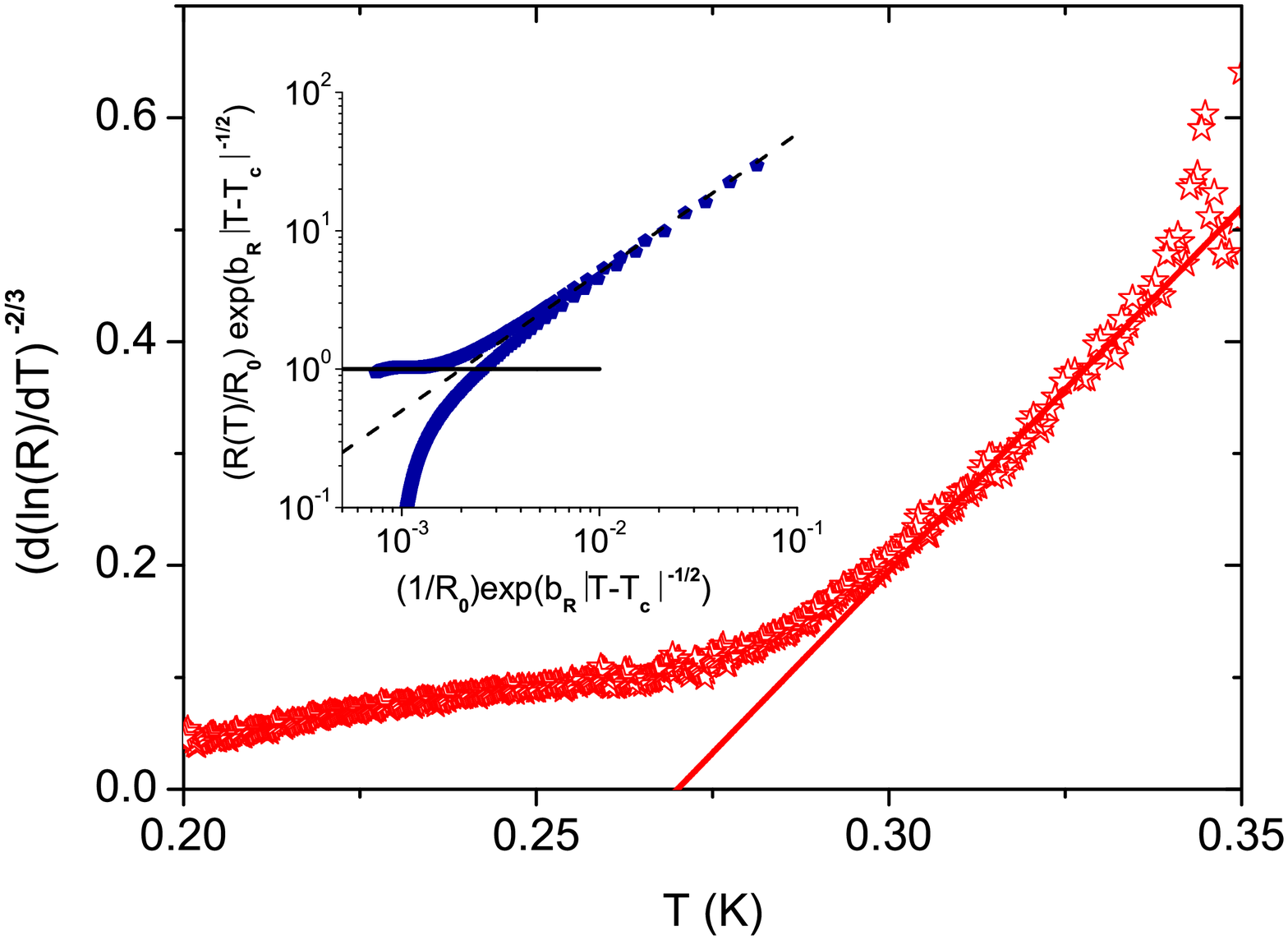}
\vspace{-1.0cm}
\caption{$\left( d\ln
R/dT\right) ^{-2/3}$ \textit{vs}. $T$ for $V_{g}=40$ V where $%
R=3/5R_{\square }$. The solid line is $\left( d\ln R/dT\right)
^{-2/3}=6.5\left( T-T_{c}\right) $ yielding the estimates $T_{c}=0.27$ K and
$\left( 2/b_{R}\right) ^{2/3}=6.5$; the inset shows $R\left( T\right) $ exp$\left( b_{R}\left\vert T-T_{c}\right\vert ^{-1/2}\right) /R_{0}$ \textit{vs}. $\left( 1/R_{0}\right) $exp$\left( b_{R}\left\vert T-T_{c}\right\vert
^{-1/2}\right) $ with $R_{0}=1.67$ k $\Omega $. The upper branch corresponds
to $T>T_{c}$ and the lower one to $T<T_{c}$. The solid line is $R\left(T,L\right) \simeq R\left( T,\infty \right) $ and the dashed one $R$ exp$\left( b_{R}\left\vert T-T_{c}\right\vert ^{-1/2}\right) /R_{0}=\left(g/R_{0}\right) $exp$\left( b_{R}\left\vert T-T_{c}\right\vert ^{-1/2}\right)
$ with $g\simeq 501\propto 1/L^{2}$.}
\label{fig1}
\end{figure}

\begin{figure}[htb]
\includegraphics[width=1.0\linewidth]{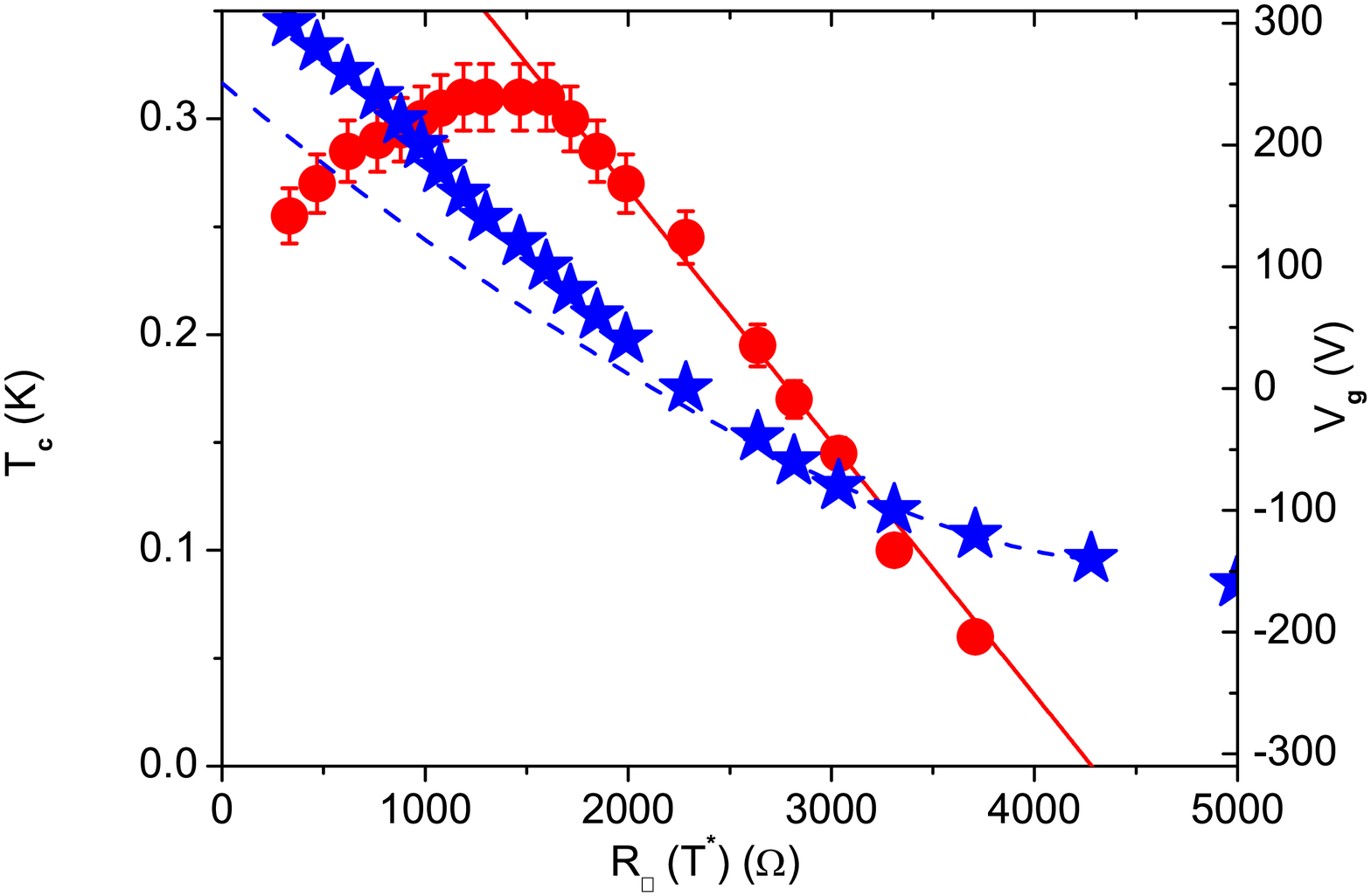}
\vspace{-1.0cm}
\caption{$T_{c}$ \textit{vs}. $R_{\square }\left( T^{\ast }\right) $ $\left( \bullet \right)$ and $V_{g}$ \textit{vs}. $R_{\square }\left( T^{\ast }\right) $ $\left( \star
\right) $ at $T^{\ast }=0.4$ K. The error bars indicate the
uncertainty in the finite size estimates of $T_{c}$. The solid line is $T_{c}$ $=1.17\times10^{-4}\Delta R(T^{\ast })$ and the dashed one $V_{g}$ $=V_{gc}+1.39\times10^{-3}$ $\Delta R^{3/2}(T^{\ast })$ with $\Delta R\left(T^{\ast }\right) =(R_{\square c}\left( T^{\ast }\right) -R_{\square }\left(
T^{\ast }\right) )$, $R_{\square c}\left( T^{\ast }\right) =4.28$ k$\Omega $
and $V_{gc}=-140$ V.}
\label{fig2}
\end{figure}

Applying this approach to the $R\left( T\right) $ data for each gate voltage
$V_{g}$ we obtain good approximations for the values of $T_{c}\left(
V_{g}\right) $, $b_{R}\left( V_{g}\right) $, and $R_{0}\left( V_{g}\right) $, in the absence of a finite size effect.  The resulting BKT-transition line is depicted in Fig. \ref{fig2}, displayed as $T_{c}$ \textit{vs}. $R_{\square }\left( T^{\ast }\right) $, the normal state
resistance at $T^{\ast }=0.4$K. We observe that it ends around $R_{\square
c}\left( T^{\ast }\right) \simeq 4.28$ k$\Omega $ where the system is
expected to undergo a 2D-QP-transition because $T_{c}$ vanishes. With
reduced $R_{\square }$ the transition temperature increases and reaches its
maximum value, $T_{cm}\simeq 0.31$ K, around $R_{\square }\left( T^{\ast
}\right) \simeq 1.35$ k$\Omega $. With further reduced resistance $T_{c}$
decreases. We also
included the gate voltage dependence of the normal state resistance since
corrections to Drude behavior ($\sigma \propto n$) have been discussed in
the literature for systems exhibiting weak localization as will be
demonstrated below.\cite{altshuler1,chakra}

\begin{figure}[htb]
\includegraphics[width=1.0\linewidth]{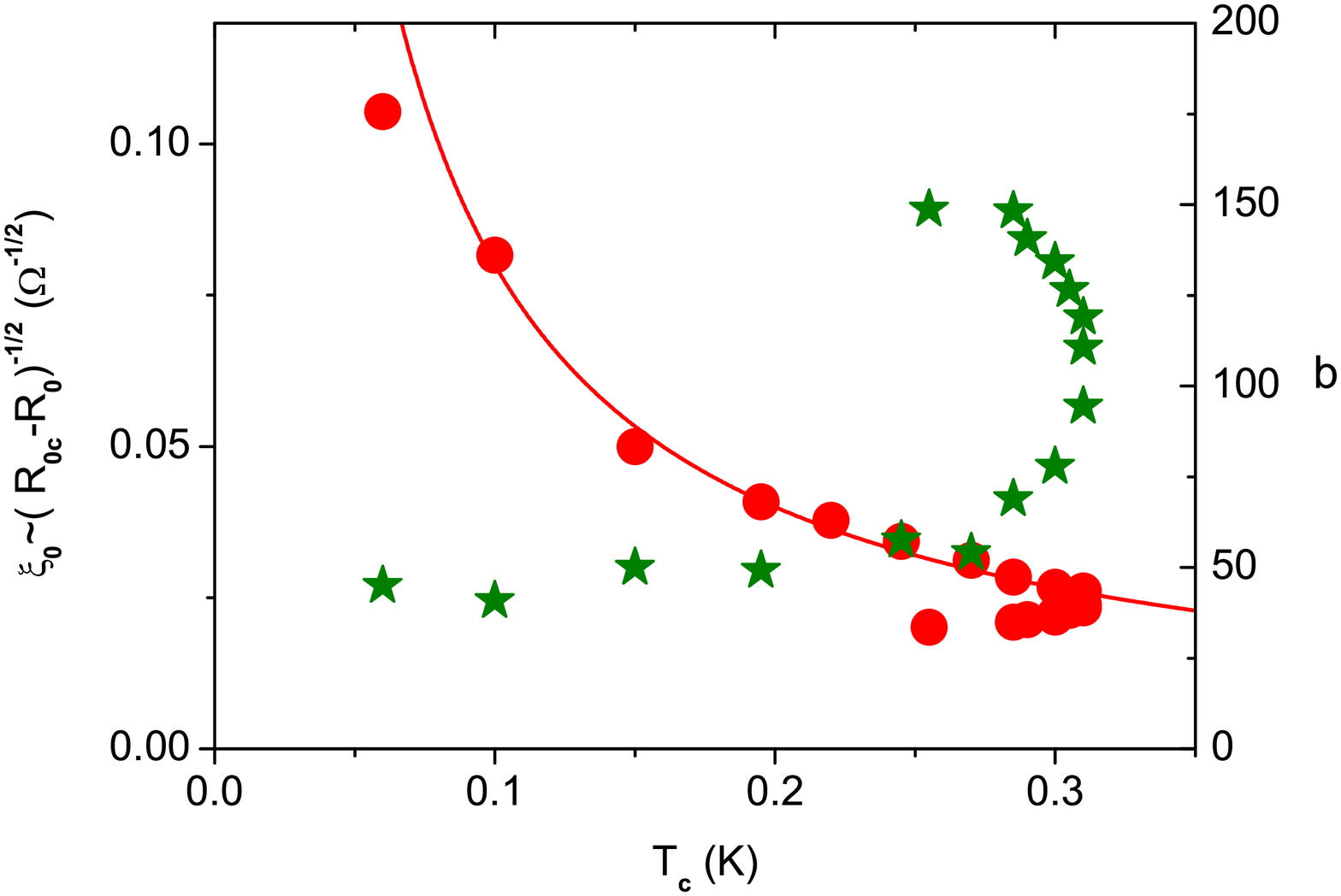}
\vspace{-1.0cm}
\caption{Vortex radius $\xi _{0}\propto \left( R_{0c}-R_{0}\right)
^{-1/2}\left( \bullet \right) $ and $b$ $\left( \bigstar \right) $ \textit{vs. } $T_{c}$ where $R=3/5R_{\square }$. The solid line is $\xi _{0}\propto \left(R_{0c}-R_{0}\right) ^{-1/2}=8\times10^{-3}/T_{c}$ with $R_{0c}=2.7$ k$\Omega$.}
\label{fig3}
\end{figure}

According to the scaling theory of quantum critical phenomena one expects
that close to the 2D-QP-transition $T_{c}$ scales as\cite{book,kim}
\begin{equation}
T_{c}\propto \delta ^{z\bar{\nu}},
\label{eq9}
\end{equation}
where $\delta $ is the appropriate scaling argument, measuring the relative
distance from criticality. $\overline{\nu }$ denotes the critical exponent
of the zero temperature correlation length $\xi \left( T=0\right) \propto
\delta ^{-\overline{\nu }}$ and $z$ the dynamic critical exponent. From Fig. \ref{fig2} it is seen that the experimental data points to the relationship
\begin{equation}
T_{c}\propto \Delta R_{\square}\left( T^{\ast }\right) \propto \Delta V_{g}^{2/3},
\label{eq10}
\end{equation}
close to quantum criticality, where $\Delta R_{\square }\left( T^{\ast
}\right) =R_{\square c}\left( T^{\ast }\right) -R_{\square }\left( T^{\ast
}\right) $. In this context it is important to emphasize that $T_{c}\propto \Delta R_{\square }\left( T^{\ast }\right) $ turns out to be nearly independent of the choice of $T^{\ast }$ around $T^{\ast }\approx 0.4$
K. So the normal state sheet resistance $R_{\square }\left( T^{\ast}\right) $ is an appropriate scaling variable in terms of $\Delta R_{\square}\left( T^{\ast }\right) $. In this case $z\overline{\nu }=1$, while if $\delta =\Delta V_{g}$, $z\overline{\nu }=2/3$. Since the measured modulation of the gate voltage induced charge density $\Delta n_{2D}$ scales in the regime of interest as \cite{caviglia}
\begin{equation}
\Delta V_{g}\propto \Delta n_{2D}\propto T_{c}^{3/2},
\label{eq11}
\end{equation}
so $z\overline{\nu }=2/3$ if $\Delta V_{g}$ or $\Delta n_{2D}$ are taken as
scaling argument $\delta $. On the other hand it is known that $\delta
\propto \Delta n_{2D}$ holds if $\left( 2+z\right) \overline{\nu }\geq 2$.\cite{fisher2} To check this inequality, given $z\overline{\nu }$, we need
an estimate of $z$. For this purpose we invoke the relation $R_{0}-R_{0c}\propto \xi _{0}^{-2}$ (Eq. (\ref{eq6})) and note that the critical amplitude of the finite temperature correlation length $\xi _{0}$
and its zero temperature counterpart should scale as $\xi _{0}\propto \xi
(T=0)\propto \delta ^{-\bar{\nu}}$, so that the scaling relation%
\begin{equation}
R_{0}-R_{0c}\propto \xi _{0}^{-2}\propto \xi ^{-2}\left( T=0\right) \propto
\delta ^{2\overline{\nu }}\propto T_{c}^{2/z},  \label{eq12}
\end{equation}
holds. Fig. \ref{fig3} depicts the $T_{c}$ dependence of the vortex core
radius $\xi _{0}\propto \xi \left( T=0\right) \propto \left(
R_{0c}-R_{0}\right) ^{-1/2}$ and $b$, which is related to the vortex energy $E_{c}$. Approaching the 2D-QP-transition we observe that the data point to $\xi \left( T=0\right) \propto 1/T_{c}$, yielding for $z$ the estimate $z\simeq 1$ so that $\bar{\nu}\simeq 2/3$ with $z\bar{\nu}\simeq 2/3$. As
these exponents satisfy the inequality $\left( 2+z\right) \overline{\nu }\geq 2$ \cite{fisher2} we identified the correct scaling argument, $\delta \propto \Delta n_{2D}\propto \Delta V_{g}$. The 2D-QP-transition is then
characterized by the scaling relations
\begin{equation}
T_{c}\propto \delta ^{z\overline{\nu }}\propto \Delta R_{\square}\left( T^{\ast
}\right) \propto \Delta V_{g}^{2/3}\propto \Delta n_{2D}^{2/3}\propto \xi
_{0}^{-1},
\label{eq13}
\end{equation}
where $\Delta R_{\square}\left( T^{\ast }\right) \propto \Delta n_{2D}^{2/3}$ reveals non-Drude behavior in the normal state. The product $z\overline{\nu }\simeq 2/3$ agrees with
that found in the electric field effect tuned 2D-QP-transition in amorphous
ultrathin bismuth films\cite{parendo} and the magnetic-field-induced 2D-QP
transition in Nb$_{0.15}$Si$_{0.85}$ films.\cite{aubin} On the contrary it
differs from the value $z\overline{\nu }\simeq 1$ that has been found in
thin NdBa$_{2}$Cu$_{3}$O$_{7}$ films using the electric-field-effect
modulation of the transition temperature.\cite{matthey} In any case our
estimates, $z\simeq 1$ and $\bar{\nu}\simeq 2/3$ point to a 2D-QP-transition which belongs to the 3D-xy universality class.\cite{book}

Fig. \ref{fig3} also depicts the $T_{c}$ dependence of $b$, which is related to the
vortex energy. Since $b$ tends to a constant in the limit $T_{c}\rightarrow 0$, Eq. (\ref{eq2}) implies $db/dT_{c}=0$ and therewith
\begin{equation}
E_{c}(T_{c})\propto k_{B}T_{c},
\label{eq13a}
\end{equation}
while the core radius diverges as
\begin{equation}
\xi _{0}\propto 1/T_{c},
\label{eq13b}
\end{equation}
in analogy to the behavior of superfluid $^{4}$He films where $T_{c}$ was tuned
by varying the film thickness.\cite{williams} A linear relationship between
the vortex core energy and $T_{c}$ was also predicted for heavily underdoped
cuprate superconductors.\cite{benfatto} Furthermore, an increase of the
vortex core radius with reduced $T_{c}$ was also observed in underdoped YBa$_{2}$Cu$_{3}$O$_{y}$\cite{sonier} and La$_{2-x}$Sr$_{x}$CuO$_{4}$. \cite{kadono} The 2D-QP-transition is then also characterized by vortices having
an infinite radius and vanishing core energy. As $T_{c}$ increases from the
2D-QP transition, the core radius shrinks, while the vortex energy increases. We also observe that the rise of $T_{c}$ is limited by a critical value of the core radius and that the maximum $T_{c}$ ($T_{cm}\simeq 0.31$
K) is distinguished by an infinite slope of both, the vortex radius and $b$.
Finally, after passing $T_{cm}$ the vortex core radius $\xi _{0}$ continues
to decrease with reduced $T_{c}$ while $b$ increases further.

\begin{figure}[htb]
\includegraphics[width=1.0\linewidth]{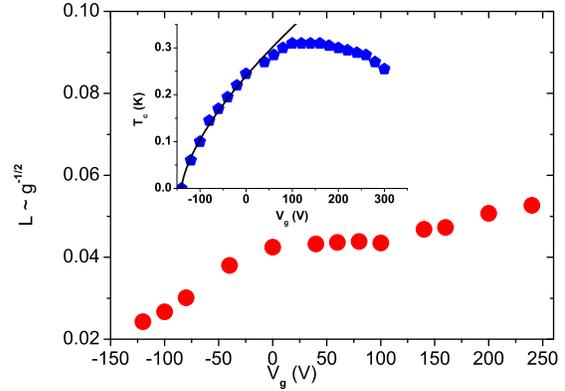}
\vspace{-1.0cm}
\caption{Gate voltage
dependence of the limiting length $L$ in terms of $\ L\propto g^{-1/2}$
\textit{vs}. $V_{g}$. The inset shows $T_{c}$
\textit{vs}. $V_{g}$. The solid line is $T_{c}=8.9\times10^{-3}(V_{g}-V_{gc})^{2/3}$ (K) indicating the leading quantum critical behavior (\ref{eq13}) with $z\bar{\nu}\simeq 2/3$.}
\label{fig4}
\end{figure}

  Next we explore the gate voltage dependence of the limiting length. Indeed,
its presence or absence allows us to discriminate between an intrinsic or
extrinsic limiting length. For this purpose we performed the finite scaling
analysis outlined in Fig. 1 for various gate voltages. In the finite size
dominated regime, $\xi >L$, the finite size scaling form (\ref{eq8}) reduces
to $R\left( T\right) $exp$\left( b_{R}\left\vert T-T_{c}\right\vert
^{-1/2}\right) /R_{0}=g(x)\propto \left( g/R_{0}\right) $exp$\left(
b_{R}\left\vert T-T_{c}\right\vert ^{-1/2}\right) $ with $g\propto 1/L^{2}$,
so $g$ probes, if there is any, the gate voltage dependence of $L$. In Fig.
4 we summarized the resulting gate voltage dependence of $L\propto g^{-1/2}$%
. It is seen that the limiting length is nearly gate voltage independent
down to $V_{g}=0$ V. This points to the presence of inhomogeneities
preventing the correlation length to grow beyond the lateral extent of the
homogeneous domains. On the contrary, for negative gate voltages $L$
decreases by approaching the QP-transition as $T_{c}$ does. The resulting
broadening of the BKT-transition with reduced $V_{g}$ and $T_{c}$ is
apparent in the temperature dependence of the sheet resistance.\cite%
{caviglia} A potential candidate for a gate voltage dependent limiting
length is the failure of cooling at very low temperatures.\cite{parendo2} In
this case the correlation length cannot grow beyond its value at the
temperature $T_{f}$ where the failure of cooling sets in. Invoking Eq. (\ref%
{eq1}) in the limit $T_{c}\rightarrow 0$ we obtain $L_{f}=\xi _{0}$exp$%
\left( 2\pi /(bT_{f}^{1/2}\right) \propto 1/T_{c}$, because $\xi _{0}\propto
1/T_{c}$ and $b$ remains finite in the limit $T_{c}\rightarrow 0$ (see Fig. %
\ref{fig3}). Contrariwise we observe in Fig. \ref{fig4} that $L$ decreases
with $T_{c}$. According to this electrostatic tuning does not change the carrier density only but the
inhomogeneity landscape as well.

 In any case, the agreement with BKT-behavior, limited by a standard finite size effect, allows us to discriminate the rounded transition from other scenarios, including strong disorder which destroys the BKT- behavior. It
also provides the basis to estimate $T_{c}$, $b_{R}$ and $R_{0}$, and with
that $b$ and $\xi _{0}\propto R_{0}^{-1/2}$ with reasonable accuracy. The
resulting BKT-transition line ends at $V_{gc}\simeq -140$ V, where $T_{c}$
vanishes and the system undergoes a 2D-QP transition. After passing this
transition $T_{c}$ increases with reduced negative gate voltage, reaches its
maximum, $T_{cm}\simeq 0.31$ K, around $V_{g}\simeq 110$ V and decreases
with further increase of the positive gate voltage. Remarkably enough, this
uncovers a close analogy to the doping dependence of $T_{c}$ in a variety of
bulk cuprate superconductors,\cite{book,tsp} where after passing the so
called underdoped limit $T_{c}$ reaches its maximum with increasing dopant
concentration. With further increase of the dopant concentration $T_{c}$
decreases and finally vanishes in the overdoped limit. This phase transition
line is thought to be a generic property of bulk cuprate superconductors.
There is, however, an essential difference. Cuprates are bulk
superconductors and the approach to the underdoped limit, where the QP
transition occurs, is associated with a 3D to 2D crossover,\cite{book,tsp}
while in the present case the system is and remains 2D, as the consistency
with BKT critical behavior reveals. Furthermore, a superconducting dome (in
the $T$ versus doping phase diagram) was also observed in bulk doped SrTiO$_{3}$ that is close to the system under study.\cite{schooley1,schooley2}

\subsection{INSULATING PHASE}

\begin{figure}[htb]
\includegraphics[width=1.0\linewidth]{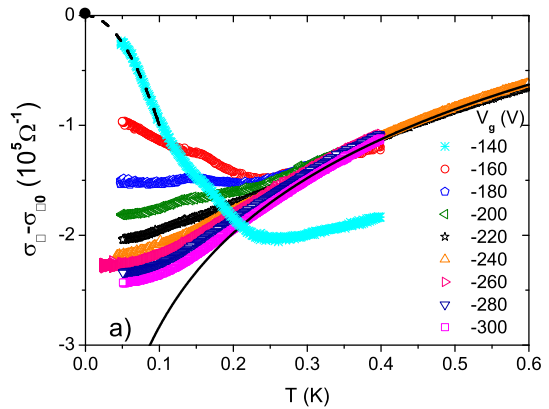}
\includegraphics[width=1.0\linewidth]{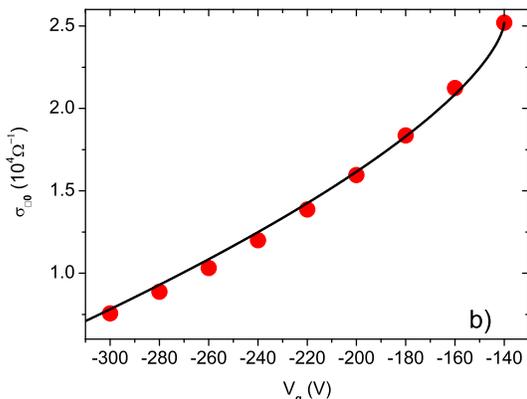}
\vspace{-1.0cm} \caption{a) $\sigma
_{\square }-\sigma _{\square 0}$ \textit{vs}. $T $ for various gate voltages
$V_{g}$. The solid line is $\sigma _{\square }-\sigma _{\square 0}=d$ ln$\left( T\right) $ ($\Omega ^{-1}$) with $d\simeq 1.23\times10^{-5}$ $\Omega^{-1}$ and $\sigma _{\square 0}\left(V_{g}\right) $\textit{\ }taken from Fig. \ref{fig5}b. The dot at the origin marks the quantum critical point and the dashed line is
Eq. (\ref{eq15}) indicating the low-temperature behavior of the sheet
conductivity at the quantum critical point. b) $\sigma
_{\square 0}$ \textit{vs}. $V_{g}$. The solid line is Eq. (\ref{eq14a}) with $
V_{gc}=-140$ V.}
\label{fig5}
\end{figure}

Supposing that the insulating phase is a weakly localized Fermi liquid the sheet conductivity
should scale as\cite{lee}
\begin{equation}
\sigma _{\square }\left( T\right) =\sigma _{\square 0}+d\ln (T),
\label{eq14}
\end{equation}
where $d=e^{2}/\left( \pi h\right) \simeq 1.23\times10^{-5}$ $\Omega ^{-1}$ is generically attributed to electron-electron interaction,\cite{blanter} while $\sigma _{\square 0}$ is expected to depend on the
gate voltage. In Fig. \ref{fig5}a we depicted $\sigma _{\square }-\sigma _{\square 0}$
\textit{vs}. $T$ for various gate voltages $V_{g}$ by adjusting $\sigma
_{\square 0}$ to achieve a data collapse at sufficiently high temperatures.
The resulting gate voltage dependence of $\sigma _{\square 0}$, consistent
with
\begin{eqnarray}
\sigma _{\square 0}\left( V_{g}\right)  &=&\sigma _{\square s}-5.9\times10^{-6}\left\vert V_{g}-V_{gc}\right\vert ^{2/3}(\Omega ^{-1}),  \nonumber \\
\sigma _{\square s} &=&2.52\times10^{-4}(\Omega ^{-1})
\label{eq14a}
\end{eqnarray}
is shown in Fig. \ref{fig5}b. An important feature of the data is the consistency
with a weakly localized Fermi liquid because the coefficient $d$ is close to
$d=e^{2}/\left( \pi h\right) $. In any case, more extended evidence for weak localization emerges from the magnetoconductivity presented below (Fig. \ref{fig9}).

\begin{figure}[htb]
\includegraphics[width=1.0\linewidth]{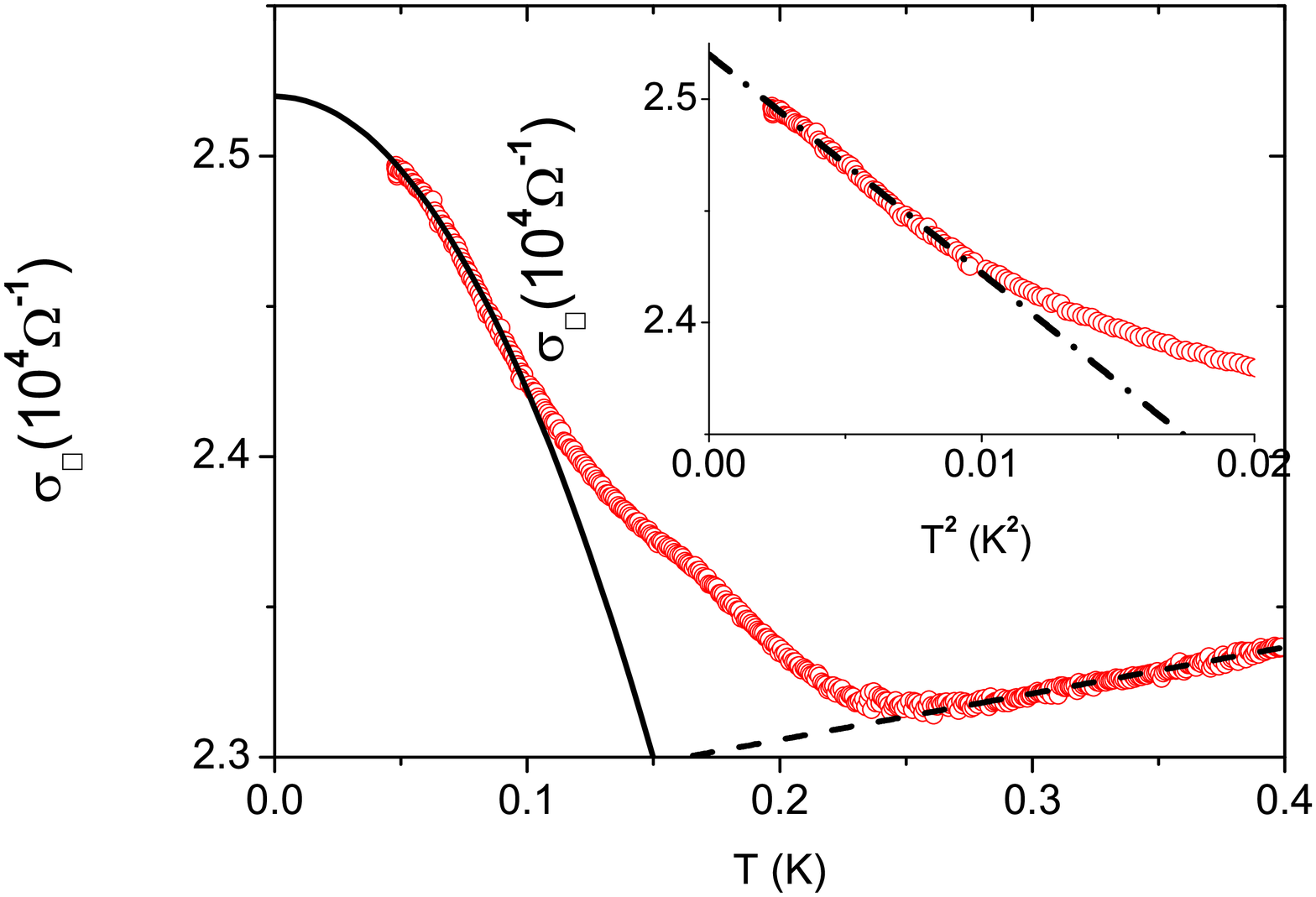}
\vspace{-1.0cm}
\caption{$\sigma _{\square }\left( V_{gc},T\right) $ \textit{vs}. $T$ and in
the inset \textit{vs}. $T^{2}$ at $V_{gc}=-140$ V. The solid line, indicating
consistency with $T^{2}$ is Eq. (\ref{eq15}), while the dashed line is $\sigma _{\square }\left( V_{gc},T\right) =2.275\times10^{-4}+1.54\times10^{-5}T$ ($\Omega ^{-1}$). The the dash-dot one in the inset is again Eq. (\ref{eq15}).}
\label{fig6}
\end{figure}

A very distinct temperature dependence of the conductivity occurs at quantum
criticality, $V_{g}=-140$ V$\simeq V_{gc}$. Indeed, the dashed line in
Fig. \ref{fig5}a and the solid one in Fig. \ref{fig6} indicate that in the limit $T\rightarrow 0$ the system tends towards a critical value. According to the plots shown in Fig. \ref{fig6} the limiting behavior is well described by
\begin{equation}
\sigma _{\square }\left( T,V_{gc}\right) =\sigma _{\square s}\left(
V_{gc}\right) -9.782\times10^{-4}T^{2}.
\label{eq15}
\end{equation}
Note that our estimate $\sigma _{\square }\left( T=0,V_{gc}\right) =\sigma
_{\square 0}\left( V_{gc}\right) =\sigma _{\square s}=2.52\times10^{-4} \Omega ^{-1}$ is comparable to the quantum unit of conductivity $4e^{2}/h\simeq 1.55\times10^{-4}$ $\Omega ^{-1}$ for electron pairs, emphasizing the importance of quantum effects.
The $T^{2}$ dependence points to Fermi liquid behavior in the regime $k_{B}T<<\hbar \omega _{D}$, $E_{F}$ where electron-electron scattering dominates. $\omega _{D}$ is the Debye frequency and $E_{F}$ denotes the
Fermi energy. At higher temperature we observe a crossover to a linear $T$
-dependent conductivity marked by the dash-dot line. Recent theories on the
conductivity of 2D Fermi liquids predict such a linear $T$-dependence.\cite{zala} From Eqs. (\ref{eq14}) and (\ref{eq14a}), describing the data in the weakly localized regime rather well, it also follows that the normal state
conductivity at $T^{\ast }=0.4$ K scales as $\sigma _{\square c}\left( T^{\ast }\right) -\sigma
_{\square }\left( T^{\ast }\right) \propto 10^{-6}\left\vert
V_{g}-V_{gc}\right\vert ^{2/3}$. Together with the empirical scaling
relation (\ref{eq13}), $\left\vert V_{g}-V_{gc}\right\vert \propto \Delta
n_{2D}$, it points to non-Drude behavior in the normal state.

Considering the temperature dependence of the sheet conductivity below quantum criticality ($V_{g}=-140$ V$\simeq V_{gc}$), Fig. \ref{fig5}a reveals at sufficiently high temperature remarkable agreement with the ln$\left(T\right) $ behavior, characteristic for weak localization. On the contrary, in the low temperature
regime and even rather deep in the insulating phase ($V_{g}=-300$ V), systematic deviations occur in terms of saturation and an upturn as quantum criticality is approached ($V_{gc}=-140$ V). Because the
conductivity of a weakly localized insulator is not expected to saturate in
the zero temperature limit \cite{marq,marq2} this behavior appears to be a
finite size effect, preventing the diverging length associated with
localization \cite{lee}, $\xi _{loc}\propto d\left\vert \ln \left( T\right)
\right\vert $, to grow beyond $L$, the limiting length already identified in
the context of the rounded BKT-transition (Fig. \ref{fig4}). In the present
case finite size scaling predicts that $\sigma _{\square }(T)$ should scale
as
\begin{equation}
\frac{\sigma _{\square }(T)-\sigma _{\square c}}{d\ln \left( T\right) }=g\left( y\right) ,\text{ }y=L/\xi _{loc}\propto L/\left( d\left\vert \ln\left( T\right) \right\vert \right) .
\label{eq16}
\end{equation}
$g(y)$ is the finite size scaling function which tends to $1$ for $y<1$. In
this case the approach to the insulating ground state can be seen, while for
$y>1$ the crossover to $g(y)\rightarrow y$ sets in and $\sigma _{\square
}(T) $ approaches the finite size dominated regime, where%
\begin{equation}
\frac{\sigma _{\square }(T)-\sigma _{\square c}}{d\ln \left( T\right) }=g_{L}/\left( d\left\vert \ln \left( T\right) \right\vert \right) ,\text{ }g_{L}\propto L
\label{eq17}
\end{equation}

\begin{figure}[htb]
\includegraphics[width=1.0\linewidth]{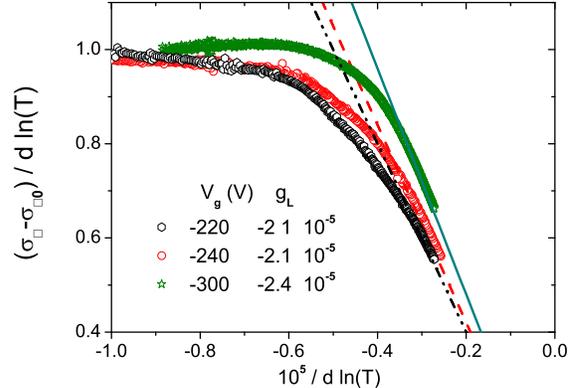}
\vspace{-1.0cm}
\caption{ $\left( \sigma
_{\square }\left( T\right) -\sigma _{\square 0}\right) /\left( d\ln
(T\right) )$ \textit{vs}. $1/\left( d\left\vert ln(T)\right\vert \right) $
at $V_{g}=-220$, $-240$ and $-300$ V with $d$ $=1.23\times10^{-5}$ $\Omega
^{-1}$. The lines correspond to Eq. (\ref{eq17}) providing
a measure for $L$ in terms of  $\left\vert g_{L}\right\vert \propto L.$}
\label{fig7}
\end{figure}

A glance at Fig. \ref{fig7}, depicting $\left( \sigma _{\square }\left( T\right)
-\sigma _{\square 0}\right) /\left( d\ln (T\right) )$ \textit{vs}. $1/\left(
d\ln (T)\right) $ at $V_{g}=-220$, $-240$ and $-300$ V, reveals that the
systematic deviations from the characteristic weak localization temperature
dependence are fully consistent with a standard finite size effect.
Accordingly, the saturation and upturns seen in Fig. \ref{fig5} at low temperatures
are attributable to a finite size effect, while in a homogeneous and
infinite system the data should collapse on the solid line in Fig. \ref{fig5}a. An essential exception is $V_{g}=-140$ V. Here the interface approaches the metallic quantum critical point (see Fig. \ref{fig5}),
metallic because the sheet conductivity remains finite, approaching $\sigma _{\square }\left(
T=0,V_{gc}\right) =\sigma _{\square 0}\left( V_{gc}\right) =\sigma _{\square
s}\simeq 2.52\times10^{-4}$ $\Omega ^{-1}$ (Eq. (\ref{eq15})) in the limit $T\rightarrow 0$.

\begin{figure}[htb]
\includegraphics[width=1.0\linewidth]{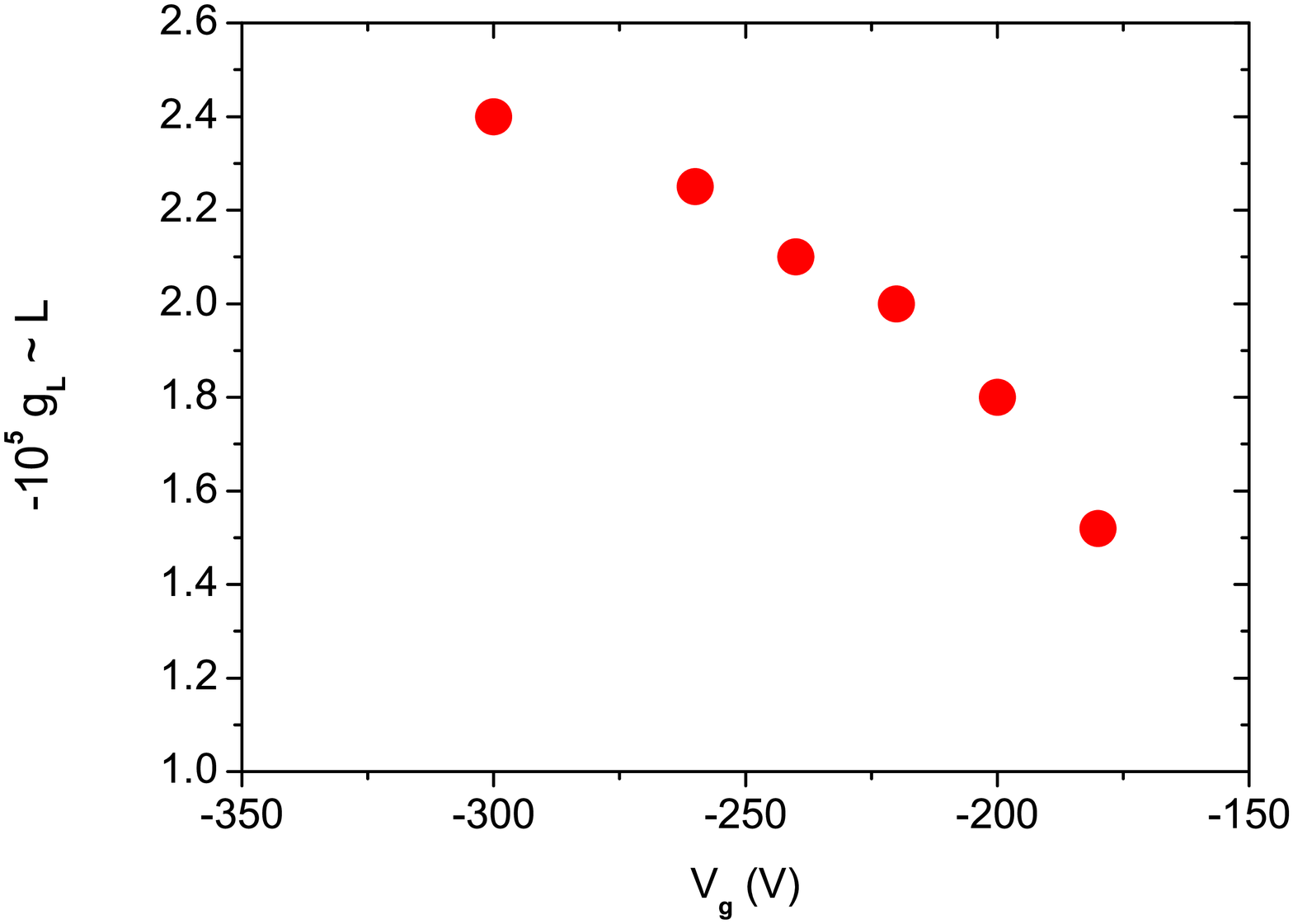}
\vspace{-1.0cm}
\caption{ $-g_{L}$ $ \propto L$ \textit{vs}. $V_{g}$ derived from finite size scaling plots as
shown in Fig. \ref{fig7} with Eq. (\ref{eq17}).}
\label{fig8}
\end{figure}

\begin{figure}[htb]
\includegraphics[width=1.0\linewidth]{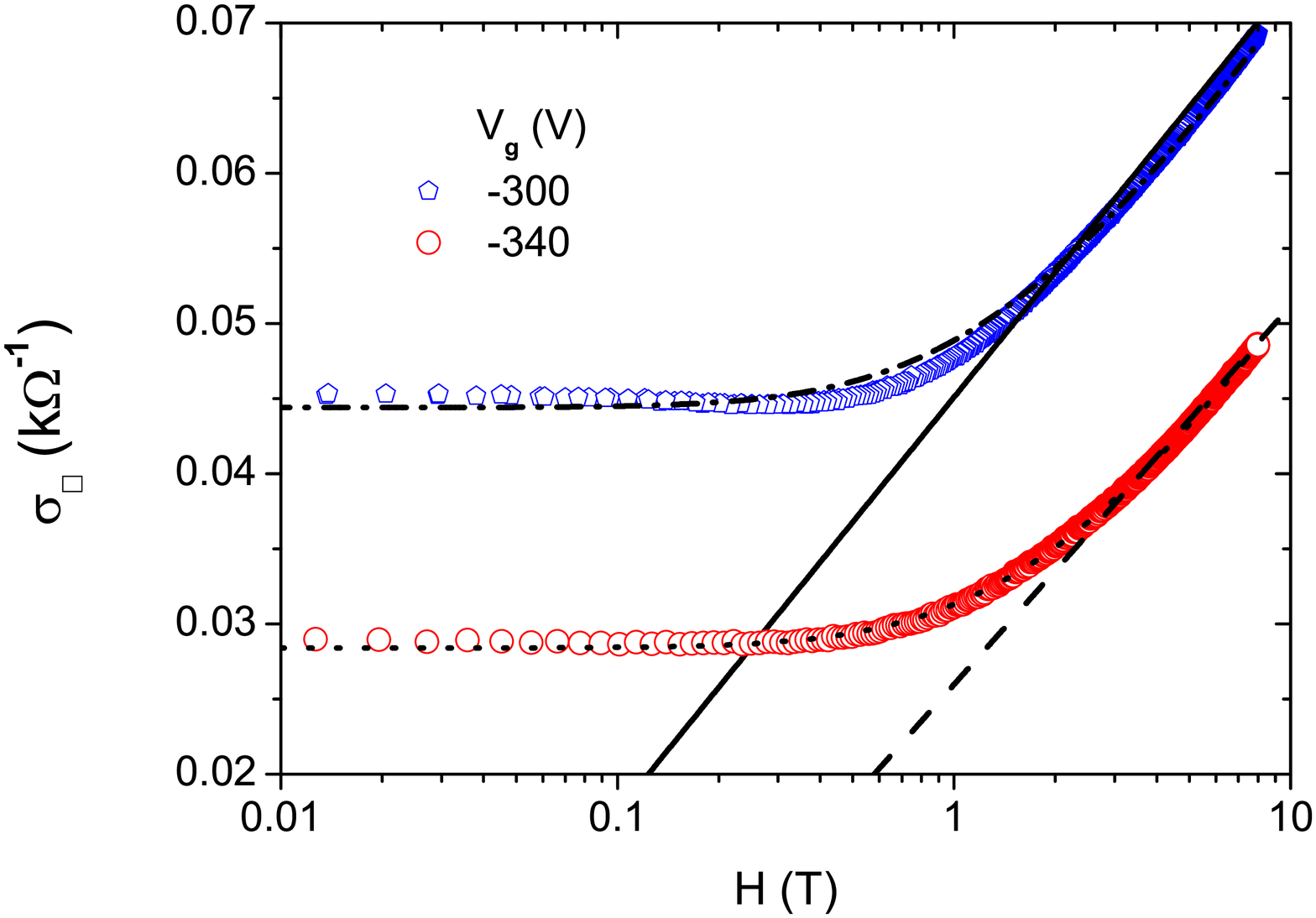}
\vspace{-1.0cm}
\caption{Magnetoconductivity $\sigma _{\square }$ \textit{vs}. $H$, applied
perpendicular to the interface, at $T=0.03$ K and $V_{g}=-300$ V and $-340$
V. The solid line is $\sigma _{\square }=4.51\times10^{-2}+1.2\cdot
10^{-2}\ln \left( H\right) $ k$\Omega ^{-1}$, the dashed one $\sigma
_{\square }=2.6\times10^{-2}+1.1\times10^{-2}\ln \left( H\right) $ k$\Omega
^{-1}$, the dotted and dash dot curves are Eq. (\ref{eq18}) with the $\sigma
_{\square 0}$, $8\pi L_{TH}^{2}/\Phi _{0}$ and $c=\alpha ^{\ast }e^{2}/(\pi
h)$ values $0.0284$ k$\Omega ^{-1}$, $2.906$ T$^{-1}$, $1.46\times10^{-5}$ $%
\Omega ^{-1}$ for $V_{g}=-340$ V and $0.044$ k$\Omega ^{-1}$, $4.068$ T$^{-1}
$, $1.46\times10^{-5}$ $\Omega ^{-1}$ at $V_{g}=-300$ V. Note that $8\pi
L_{TH}^{2}/\Phi _{0}=$ $2.906$ T$^{-1}$ corresponds to $L_{TH}\simeq
1.55\times10^{-6}$cm whereupon $L_{H}=L_{TH}$ at $H=1.37$ T.}
\label{fig9}
\end{figure}

Fig. 7 also reveals that the limiting length, $L\propto $ $g_{L}$, depends
in the insulating phase on the gate-voltage as well. The resulting
dependence, $L\left( V_{g}\right) \propto $ $g_{L}\left( V_{g}\right) $, is
shown in Fig. \ref{fig8}. In analogy to the limiting length associated with
the BKT-transition (Fig. \ref{fig4}) it decreases by approaching quantum
criticality at $V_{g}\simeq -140$ V. As a reduction of $L$ enhances
deviations from the asymptotic behavior this feature accounts for the
saturation and upturns seen in Fig. \ref{fig5}a. Supposing that the limiting
length is set by the failure of cooling below the temperature $T_{f}$ then $L
$ is set by $L_{f}=\xi _{loc}\left( T_{f}\right) \propto d\left\vert \ln
\left( T_{f}\right) \right\vert $ and with that independent of the gate voltage,
in disagreement with Fig. \ref{fig8}. Accordingly, in analogy to the BKT-transition, the limiting length appears to be attributable to a electrostatic mediated change of the inhomogeneity landscape.

Direct experimental evidence for a limiting length emerges from the work of
Ilani \textit{et al}.\cite{ilani} A single electron transistor was used as
a local electrostatic probe to study the underlying spatial structure of the
metal-insulator transition in two dimensions. The measurements show that as
the transition is approached from the metallic side, a new phase emerges
that consists of weakly coupled fragments of the two-dimensional system.
These fragments consist of localized charge that coexists with the
surrounding metallic phase. As the density is lowered into the insulating
phase, the number of fragments increases on account of the disappearing
metallic phase. The measurements suggest that the metal-insulator transition
is a result of the microscopic restructuring that occurs in the system.  On
the other hand, we have seen that the limiting length associated with the
resulting inhomogeneities depends on the gate voltage (see Figs. \ref{fig4}
and \ref{fig8}).

 Further evidence for a weakly localized insulating phase stems from the
observed negative magnetoresistance.\cite{caviglia} An applied magnetic
field leads to a new length given by the size of the first Landau orbit, or
magnetic length, $L_{H}=$ $\left( \Phi _{0}/\left( 2\pi H\right) \right)
^{1/2}$, which decreases with growing field strength. Once its size becomes
comparable to the dephasing length $L_{Th}$ (distance between inelastic
collisions)\cite{thouless} weak localization is suppressed. In D=2 the
following formula for the magnetoconductivity was obtained:\cite{altshuler,hikami}

\begin{equation}
\sigma _{\square }=\sigma _{\square 0}+c\left[ \psi \left( 1/2+1/x\right)
+\ln \left( x\right) \right] ,\text{ }c=\frac{\alpha ^{\ast }e^{2}}{\pi h},
\label{eq18}
\end{equation}
where $\psi $ denotes the digamma function, $\alpha ^{\ast }$ is a constant
of the order of unity,\cite{hikami} and
\begin{equation}
x=\frac{8\pi L_{Th}^{2}H}{\Phi _{0}}.
\label{eq19}
\end{equation}
In the limit $x>>1$ it reduces to
\begin{equation}
\sigma _{\square }=\sigma _{\square 0}+\frac{\alpha ^{\ast }e^{2}}{\pi h}
\left[ -1.96+\ln \left( x\right) \right],
\label{eq20}
\end{equation}
while in the limit $x\rightarrow 0$
\begin{equation}
\sigma _{\square }-\sigma _{\square 0}\propto H^{2},
\label{eq21}
\end{equation}
holds. Here
\begin{equation}
\frac{d\sigma _{\square }}{d\ln \left( H\right) }=\frac{\alpha ^{\ast }e^{2}}{\pi h}\simeq \alpha ^{\ast }1.24\times10^{-5}\Omega ^{-1},
\label{eq22}
\end{equation}
applies. In Fig. \ref{fig9} we compare the experimental data with the theoretical predictions. The data agrees reasonably well with the characteristic weak localization behavior (Eq. (\ref{eq18})), while the
asymptotic ln$(H)$ behavior (Eq. (\ref{eq20})) is not fully attained. The resulting estimates
for $d\sigma _{\square }/d\ln \left( H\right) $ are close to $e^{2}/\left(
\pi h\right) \simeq 1.24\times10^{-5}\Omega ^{-1}$ and consistent with the zero field
temperature dependence of the sheet conductivity, $\sigma _{\square }\left(
T\right) =\sigma _{\square 0}+d\ln (T)$, with $d=e^{2}/\left( \pi h\right) $
(see Fig. \ref{fig5}a). An analogous treatment
of the magnetoresistance data of a non superconducting sample of Brinkman \textit{et al}.\cite{brinkman} yields $8\pi L_{Th}^{2}/\Phi _{0}=2.862$ T$^{-1}$ and $c=\alpha ^{\ast }e^{2}/\pi h=4.8\times10^{-5}$ $\Omega ^{-1}$, so $c$ adopts in 'superconducting' and 'non superconducting' samples substantially different values. In any case, our analysis of the magnetoconductivity uncovers a weakly localized insulating phase, consistent with the ln($T$) temperature dependence of the zero field counterpart at sufficiently high temperatures, and non-Drude behavior in the normal state.

\subsection{QUANTUM PHASE TRANSITION}

\begin{figure}[htb]
\includegraphics[width=1.0\linewidth]{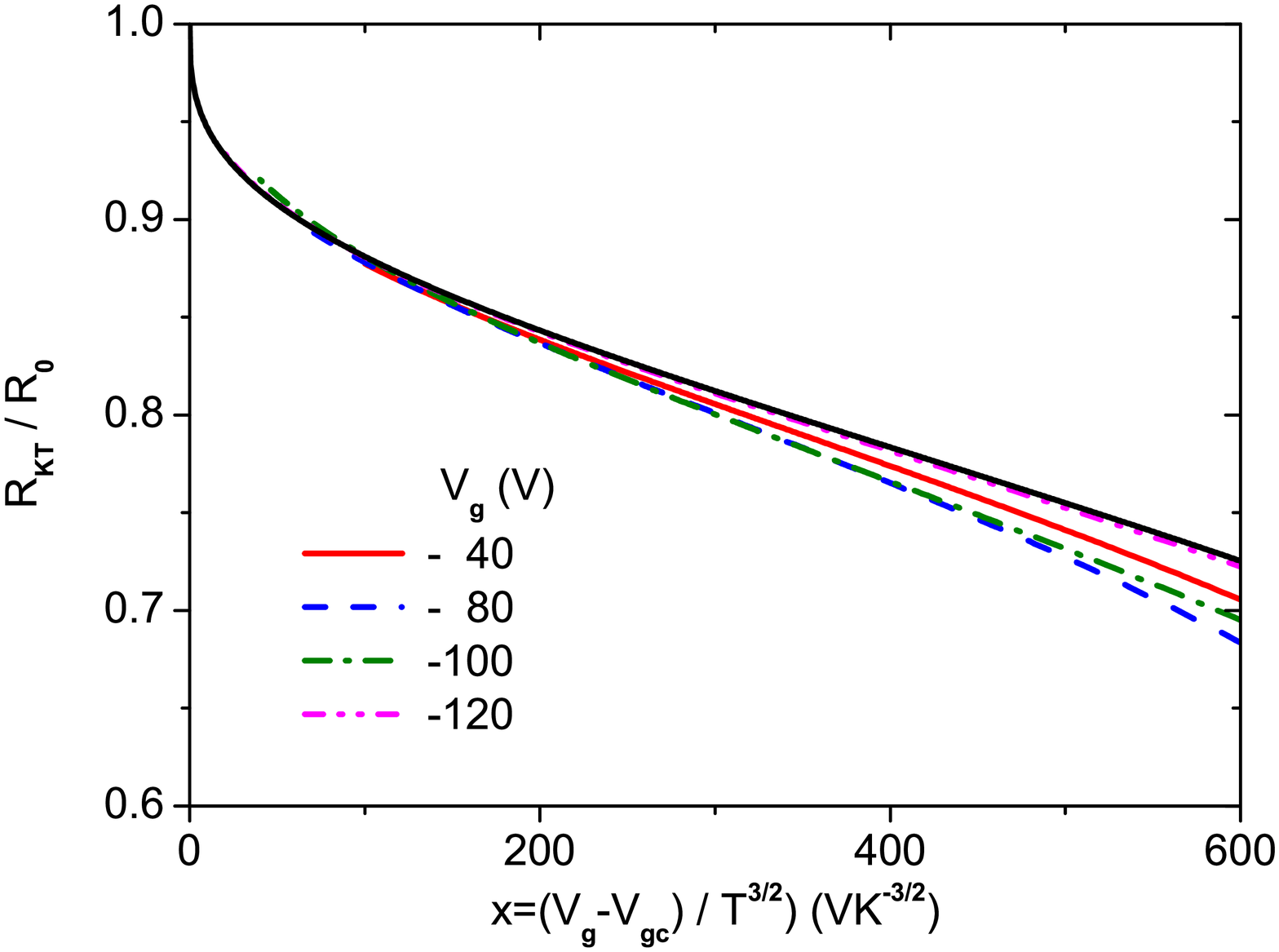}
\vspace{-1.0cm}
\caption{$R_{KT}\left( V_{g},T\right) /R_{0}\left( V_{g}\right) $ \textit{vs}. $\left( V_{g}-V_{gc}\right) /T^{3/2}$ for various $V_{g}$'s. The solid line is Eq. (\ref{eq24a}) with $\widetilde{a}=0.0248$ and $\widetilde{b}=0.0081$ in K$^{3/2}$V$^{-1}$.}
\label{fig10}
\end{figure}

Traditionally the interpretation of experimental data taken close to the
2D-QP-transition was based on the quantum scaling relation\cite{mpaf,goldman,goldman2}
\begin{equation}
R_{\square }\left( T,\delta \right) =R_{\square s}G\left( x\right) ,\text{ }x=c\delta /T^{1/z\bar{\nu}}.
\label{eq23}
\end{equation}
$G\left( x\right) $ is a scaling function of its argument and $G\left(0\right) =1$, so at quantum criticality the system is metallic with sheet resistance $R_{\square s}$. The BKT-line is then fixed by $x_{c}=c\delta
/T_{c}^{1/z\bar{\nu}}$, whereby $T_{c}$ vanishes as $T_{c}\propto \delta ^{z\bar{\nu}}\propto \xi ^{-z}\left( T=0\right) $ (Eq. (\ref{eq13})). $c$ is a nonuniversal parameter and $\delta $ the appropriate scaling argument,
measuring the relative distance from criticality. This scaling form follows
by noting that the divergence of $\xi \left( T=0\right) \propto \delta ^{-\bar{\nu}}$ is at finite temperature cut off by a length $L_{T}$, which is determined by the temperature: $L_{T}\propto T^{-1/z}$. Thus $G\left(
x\right) $ is a finite size scaling function because $x\propto \left(L_{T}/\xi \left( T=0\right) \right) ^{1/\bar{\nu}}\propto \delta /T^{1/z\bar{\nu}}$. The data for $R_{\square }\left( T,\delta \right) $ plotted \textit{vs}. $\delta /T^{1/z\bar{\nu}}$ should then collapse onto two branches joining at $R_{\square s}$. The lower branch stems from the superconducting and the upper one from the insulating phase. To explore the consistency with the
critical BKT-behavior we note that in the limit $T_{c}\rightarrow 0$ the relation
\begin{eqnarray}
\frac{R_{KT}\left( T,V_{g}\right) }{R_{0}\left( V_{g}\right) } &=&\exp
\left( -\frac{b_{R}\left( V_{g}\right) }{\left( T-T_{c}\left( V_{g}\right)
\right) ^{1/2}}\right)   \nonumber \\
&=&G\left( \frac{V_{g}-V_{gc}}{T^{1/z\bar{\nu}}}\right) ,
\label{eq24}
\end{eqnarray}
should apply. Indeed, the BKT-scaling form of the resistance applies for any
$T\gtrsim T_{c}$\ because the universal critical behavior close to $T_{c}$
is entirely classical. \cite{vojta} On the contrary $T_{c}$, $b_{R}$ and the
critical amplitude $R_{0}$ are non-universal quantities which depend on the
tuning parameter. Furthermore, they are renormalized by quantum
fluctuations. In any case, the data plotted as $R_{KT}\left( T,V_{g}\right)
/R_{0}\left( V_{g}\right) $ \textit{vs}. $\left( V_{g}-V_{gc}\right) /T^{1/z\bar{\nu}}$ should collapse on a single curve and approach one close to quantum criticality. In Fig. \ref{fig10} we depicted this scaling plot, derived from $R_{0}\left( V_{g}\right) $, $b_{R}\left( V_{g}\right) $ and $T_{c}\left(V_{g}\right) $ for $z\overline{\nu }=2/3$. Apparently, the flow to the quantum critical point is well confirmed. Furthermore, noting that close to the QP-transition $b_{R}\left( V_{g}\right) \propto T_{c}^{1/2}$, because $b_{R}=4\pi T_{c}^{1/2}/b$ (Eq. (\ref{eq3})) and $b\simeq const.$ (see Fig. \ref{fig3}), the vortex core energy scales as $E_{c}\propto
k_{B}T_{c}$ (Eq. (\ref{eq13a})), the scaling function adopts with $z\bar{\nu}=2/3$ and $T_{c}$ $\propto \left( V_{g}-V_{gc}\right) ^{2/3}$ (Eq. (\ref{eq13})) the form
\begin{equation}
G\left( x\right) \simeq \exp (-\widetilde{a}x^{1/3}/(1-\widetilde{b}
x^{2/3})^{1/2}),\text{ }x=\frac{V_{g}-V_{gc}}{T^{3/2}},
\label{eq24a}
\end{equation}
shown by the solid line in Fig. \ref{fig10}. Since $V_{g}-V_{gc}\propto T_{c}^{1/z\bar{\nu}}\propto T_{c}^{3/2}$ and $b_{R}=4\pi T_{c}^{1/2}/b$ is related to the vortex energy in terms of $b$ (Eq. (\ref{eq3})), the scaling function is
controlled by the BKT-line and the vortex core energy, while the vortex
core radius enters the prefactor via
$R_{0}\left( V_{g}\right)-R_{0c}\propto \xi _{0}^{-2}\propto \xi ^{-2}\left( T=0\right) \propto
T_{c}^{2}$ (Eq. (\ref{eq12})). Noting that $\widetilde{a}=4\pi a^{1/2}/b$ and $\widetilde{b}=a$, where $a$ is given in terms of $T_{c}=a(V_{g}-V_{gc})^{2/3}$ (K) with $a\simeq 8.9\cdot 10^{-3}$ (see Fig. \ref{fig4}) and $b\simeq 50$ (Fig. \ref{fig3}) we obtain $\widetilde{a}\simeq 0.0237$ and $\widetilde{b}=a\simeq 0.0089$, in reasonable agreement with the fit parameters yielding the solid line in Fig. \ref{fig10}. This uncovers the consistency and reliability of our estimates along the BKT-line. In this context it should be kept in mind that our analysis of the
insulating state is limited by the finite size effect, preventing to approach the zero temperature regime.

On the contrary, in the insulating phase we observed that the sheet
conductivity scales according to Eqs. (\ref{eq14}) and (\ref{eq14a}) as
\begin{equation}
\frac{\sigma _{\square }\left( T,V_{g}\right) }{\sigma _{\square s}}=1-\frac{5.9\times10^{-6}}{\sigma _{\square s}}\left\vert V_{g}-V_{gc}\right\vert^{2/3}+\frac{d}{\sigma _{\square s}}\ln (T).
\label{eq25}
\end{equation}
which is incompatible with the standard scaling form (\ref{eq23}). Indeed,
it involves two independent lengths. $\xi _{ld}\propto 1/\left\vert \ln
(T)\right\vert $, the diverging length associated with localization \cite{lee} and $\xi \left( T=0\right) \propto \left\vert \Delta V_{g}\right\vert^{-2/3}$, the zero temperature correlation length (Eq. (\ref{eq13})). In this context it should be kept in mind that our analysis of the
insulating state does not extend to zero temperature because Eq. (\ref{eq25}) applies at finite temperatures only. As $T$ is reduced further the question of what happens in the insulating phase remains.

  To complete the BKT- and 2D-QP-transition scenario measurements of the
magnetic penetration depth, $\lambda \left( T\right) $, would be required.
At the BKT- transition $T_{c}$ and $\lambda \left( T\right) $ are related by%
\begin{equation}
\rho _{s}\left( T_{c}\right) =\frac{d\Phi _{0}^{2}}{16\pi ^{3}\lambda
^{2}\left( T_{c}\right) }=\frac{2}{\pi }k_{B}T_{c},
\label{eq26}
\end{equation}
while $\rho _{s}\left( T\right) =0$ above $T_{c}$. $\rho _{s}$ is the 2D
superfluid density and $d$ is the thickness of the superconducting sheet
\cite{nelson}. The presence or absence of the resulting Nelson-Kosterlitz
jump would then allow to discriminate experimentally between weak and strong
disorder. In this context we note that there is the Harris criterion, \cite{harris} which states that short-range correlated and uncorrelated disorder
is irrelevant at the unperturbed critical point, provided that $2-D\nu <0,$where $D$ is the dimensionality of the system and $\nu $ the critical exponent of the finite temperature correlation length. With $D=2$ and $\nu
=\infty $, appropriate for the BKT-transition,\cite{kosterlitz} any
rounding of the jump should then be attributable to the finite size effect
stemming from the limiting length $L$. Furthermore, there is the quantum
counterpart of the Nelson-Kosterlitz relation, stating that
\begin{equation}
\frac{d}{\lambda ^{2}\left( T=0\right) }=\frac{16\pi ^{3}k_{B}T_{c}Q_{2}}{%
\Phi _{0}^{2}},
\label{eq27}
\end{equation}
close to the 2D-QP transition.\cite{book,tsben,kim} $Q_{2}$ is a
dimensionless critical amplitude bounded by\cite{herbut}
\begin{equation}
\frac{2}{\pi }<Q_{2}<1.11.
\label{eq27a}
\end{equation}
The lower bound corresponds to the BKT-line, $d/\lambda ^{2}\left(
T=0\right) \simeq 1.03T_{c}$ with $d$, $\lambda $ in $cm$ and $T_{c}$ in K.
Below this line the superfluid order would become unstable to unbinding of
vortices. The upper bound, corresponds to $d/\lambda ^{2}\left( T=0\right)
\simeq 1.61T_{c}$ and the transition at $T=0$ belongs to the
BKT-universality class and consequently at $T_{c}$ the superfluid density
exhibits the universal discontinuity. Given our evidence for a (2+1)-xy QP
transition, quantum fluctuations are present and expected to reduce $Q_{2}$ from its maximum value, while the finite-temperature transition remains again in the BKT universality class.\cite{herbut} Correspondingly,
measurements of the temperature and gate voltage dependence of the
superfluid density would be desirable to explore the observed BKT-behavior,
weak localization and Fermi liquid features further.

\section{SUMMARY AND DISCUSSION}

In summary, we have shown that the electrostatically tuned phase transition
line at the LaAlO$_{3}$/SrTiO$_{3}$ interface, observed by Caviglia \textit{et al}.,\cite{caviglia} is consistent with a BKT-line ending at a 2D-quantum
critical point with critical exponents $z\simeq 1$ and $\overline{\nu }\simeq 2/3$, so the universality class of the transition appears to be that
of the classical 3D-xy model. We have shown that the rounding of the
BKT-transition line and the saturation of the sheet conductivity close to
the QP-transition are remarkably consistent with a gate voltage dependent
finite size effect. According to this, electrostatic tuning does not change
the carrier density only but the inhomogeneity landscape as well. Taking the resulting finite size effect into account we provided consistent evidence  for a weakly localized insulator separated from the superconducting phase by a
metallic ground state at quantum criticality. Consistent with the non-Drude
behavior in the normal state, characteristics of weak localization have been
identified in both, the temperature and magnetic field dependence of the
conductivity. The conductivity along the BKT-transition line was found to
agree with the standard scaling form of quantum critical phenomena, while in
the weakly localized insulating phase it appears to fail in the accessible
temperature regime. As in the quantum scaling approach the scaling function
is unknown we obtained its form in the superconducting phase. It is controlled
by the BKT-phase transition line and the vortex energy. In addition we explored the $T_{c}$ dependence of the
vortex core radius and the vortex energy. As the nature of the metallic
ground state at quantum criticality is concerned, the limiting $T^{2}$
dependence of the sheet conductivity points to Fermi liquid behavior,
consistent with the evidence for weak localization in the insulating phase
and non-Drude behavior in the normal state. In conclusion we have shown that the appearance of metallicity at the interface between insulators, a wonderful example of how subtle changes in the structure of these systems
can lead to fundamental changes in physical properties, is a source of rich
physics in two dimensions.

This work was partially supported by the Swiss National Science Foundation
through the National Center of Competence in Research, and ``Materials with
Novel Electronic Properties, MaNEP'' and Division II.

\end{document}